\documentclass[pra,showpacs,showkeys,twocolumn]{revtex4}
\usepackage{graphicx,amsmath,amsfonts,amssymb}
\usepackage{times}
\begin{document}
\title{Manipulation of tripartite-to-bipartite entanglement localization under quantum noises and its application to entanglement distribution}
\author{Xin-Wen Wang,$^{1,2,}$\footnote{xwwang@mail.bnu.edu.cn} Shi-Qing Tang,$^{2}$ Ji-Bing Yuan,$^{2}$ and Le-Man Kuang$^{2,}$\footnote{lmkuang@hunnu.edu.cn}}
 \affiliation{$^1$Department of Physics and Electronic Information Science, Hengyang Normal University, Hengyang 421002, China\\
  $^2$ Key Laboratory of Low-Dimensional Quantum Structures and Quantum Control of Ministry of
  Education, and Department of Physics, Hunan Normal University, Changsha 410081, China}

\begin{abstract}
This paper is to investigate the effects of quantum noises on
entanglement localization by taking an example of reducing a
three-qubit Greenberger-Horne-Zeilinger (GHZ) state to a two-qubit
entangled state. We consider, respectively, two types of quantum
decoherence, i.e., amplitude-damping and depolarizing decoherence,
and explore the best von Neumann measurements on one of three qubits
of the triple GHZ state for making the amount of entanglement of the
collapsed bipartite state be as large as possible. The results
indicate that different noises have different impacts on
entanglement localization, and that the optimal strategy for
reducing a three-qubit GHZ state to a two-qubit one via local
measurements and classical communications in the amplitude-damping
case is different from that in the noise-free case. We also show
that the idea of entanglement localization could be utilized to
improve the quality of bipartite entanglement distributing through
amplitude-damping channels. These findings might shed a new light on
entanglement manipulations and transformations.
\end{abstract}

\pacs{03.67.Bg, 03.67.Pp, 03.65.Yz, 03.67.Mn}

\keywords{Entanglement distribution, entanglement localization, GHZ
state, decoherence}

\maketitle

\section{Introduction}

Establishment of entanglement among distant parties is a
prerequisite for implementing lots of remote quantum-information
processing tasks \cite{81RMP865, 84RMP777}. In situations of
practical interest, most of these scenarios involve many parties,
and the specific subsets which will carry out quantum communications
are not known when the entangled resources are generated and
distributed among all of the parties. Particularly, different nodes
in a quantum network are usually connected by multipartite entangled
states \cite{3NP256,453N1023}, and the two-party quantum
communication protocols between any two possible parities are not
set in advance. For accomplishing two-party quantum communications,
they need to previously establish bipartite entanglement between
them via the help of other parties \cite{77PRA022308}. It is hence
interesting to search efficient ways to extract entangled states
with fewer particles (e.g., two particles) from multiparticle
entangled states.

Many theoretical works study, as a method of establishing
entanglement between two of many parties who previously share a
multipartite entangled state, a reduction the multipartite entangled
state to a bipartite entangled state via local measurements assisted
by classical communications. Such a paradigm of localizing bipartite
entanglement is related to the notions of entanglement-of-assistance
\cite{EOA,3QIC64}, localizable-entanglement
\cite{92PRL027901,71PRA042306}, and entanglement-of-collaboration
\cite{74PRA052307}. They quantify the maximal average amount of
entanglement of two parties that can be extracted from a
multipartite entangled state via (local) measurements and different
ways of classical communications. From the practical point of view,
however, it may be more important to maximize the entanglement
between the chosen two parties for specific events, where the
desired measurement outcomes of other parties are gotten, as shown
in this paper.

The idea of entanglement localization works perfectly for ideally
isolated systems. In practice, however, no system can be completely
isolated from surroundings \cite{82RMP1155}, and the system will
experience decoherence because of the interaction with environment.
Multipartite entanglement, which holds much richer quantum
correlations than bipartite entanglement, is known to be very
fragile to decoherence and to display subtle decay features
\cite{100PRL080501,78PRA064301,81PRA064304,45JPB225503}, especially
when an entangled multiparticle state is distributed into several
distant recipients \cite{453N1023,76RPP096001}. Then the
conventional entanglement localization strategies may achieves no
longer the optimization. Therefore, it is important to understand
and optimize techniques to realize effective entanglement
localization in the face of noise and decoherence.

In this paper, we investigate the effects of quantum noises on the
tripartite-to-bipartite entanglement localization and the optimal
single-particle measurement strategy for reducing a three-qubit
Greenberger-Horne-Zeilinger (GHZ) state \cite{GHZ} to a two-qubit
entangled state. We show that the amplitude and depolarizing noises
have different impacts on entanglement localization, and that the
best von Neumann measurement on one of three qubits of a triple GHZ
state for extracting a two-qubit entangled state in the
amplitude-damping environment is different from that in the
noise-free and depolarizing cases. These results indicate that when
considering the amplitude-damping decoherence, the three parties who
previously share a three-qubit GHZ state should take different
entanglement localization strategy from that in the ideal case, for
increasing the amount of entanglement of the final two-qubit
entangled state. In addition, we also demonstrate that the idea of
entanglement localization could be utilized to improve the quality
of bipartite entanglement distribution.

The paper is organized as follows. In section II, we describe the
process of entanglement localization from a three-qubit GHZ state to
a two-qubit entangled state and give the optimal measurement basis
of anyone of the three qubits. In section III, we show how can the
idea of entanglement localization boost the quality of bipartite
entanglement distribution. Concluding remarks are given in section
IV.

\section{Tripartite-to-bipartite entanglement localization under quantum noises}

GHZ states, typical multipartite maximally entangled states, are
usually employed for entanglement distribution among different nodes
of a quantum network \cite{57PRA822,103PRL020501}, due to the fact
that they can be used to implement numerous quantum information
protocols \cite{81RMP865,84RMP777}. On the other hand, the
characteristics of a GHZ state with many bodies could be usually
obtained by straightforwardly generalizing that of tripartite GHZ
states \cite{155PLA441,59PRA1829,69PRA052307,8IJQI1301}. In
consequence of these facts, we here focus on entanglement
localization of tripartite GHZ states. Considering the case that
Alice, Bob, and Charlie, staying far away from each other,
previously share a three-qubit GHZ state
\begin{equation}
|\psi\rangle^{(123)}=\frac{1}{\sqrt{2}}(|000\rangle+|111\rangle)_{123},
\label{GHZ}
\end{equation}
where $\{|0\rangle,|1\rangle\}$ is the computational basis of a
qubit. Qubits 1, 2, and 3 are in the labs of Alice, Bob, and
Charlie, respectively. Now two of them, e.g., Alice and Bob, want to
implement private quantum communication with the existing quantum
resource, the GHZ-type entangled state. To this end, they need to
first establish bipartite entanglement between them through the
assistance of the third party, Charlie. The easiest and robust
method is that Charlie performs a local measurement on qubit 3 and
broadcasts the outcome, this is so called entanglement localization
\cite{92PRL027901,71PRA042306}. Ideally, that is, in the noise-free
case, the best measurement that Charlie should adopt is a projective
measurement with basis
$\left\{|\pm\rangle=(|0\rangle\pm|1\rangle)/\sqrt{2}\right\}$,
because Alice and Bob can attain a maximally entangled state, the
Bell state
$|B_+\rangle^{(12)}=(|00\rangle+|11\rangle)_{12}/\sqrt{2}$ or
$|B_-\rangle^{(12)}=(|00\rangle-|11\rangle)_{12}/\sqrt{2}$, for each
possible measurement outcome, $|+\rangle$ or $|-\rangle$. As a
matter of fact, the average amount of entanglement between Alice and
Bob is one being equivalent to the localizable-entanglement allowed
in this case \cite{92PRL027901,71PRA042306}. The procedure of the
entanglement localization is schematically sketched in Fig.~1 (a).

\begin{figure}
\includegraphics[width=7cm,height=9cm]{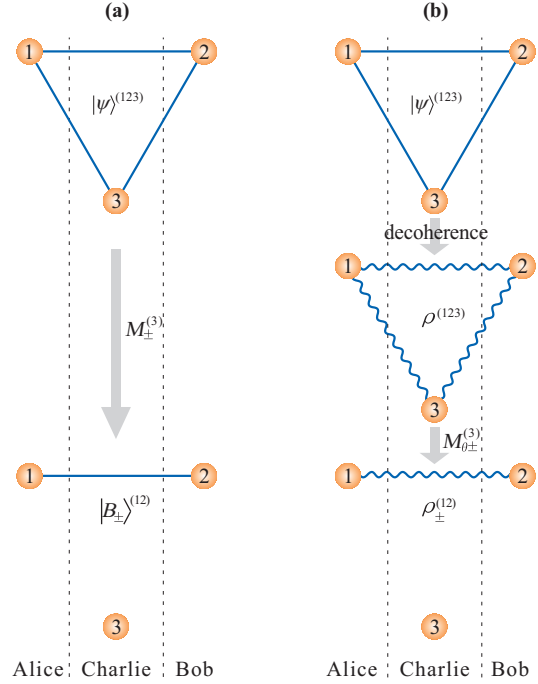}
\caption{(Color online) Sketch map of entanglement localization for
the initial three-qubit GHZ state. Diagram \textbf{(a)} describes
the ideal case where the system is isolated perfectly from its
surroundings and does not suffer from decoherence; diagram
\textbf{(b)} describes the case where each qubit undergoes
decoherence before the performance of entanglement localization
protocol. The qubits that are linked by straight lines are in a
maximally entangled pure state, and that are linked by wave lines
are in a mixed state. $M^{(3)}_{\theta\pm}$ ($M^{(3)}_{\pm}$)
denotes a von Neumann measurement on qubit 3 with projectors
$M^{(3)}_{\theta+}=|+_{\theta}\rangle_3\langle+_{\theta}|$
($M^{(3)}_+=|+\rangle_3\langle+|$) and
$M^{(3)}_{\theta-}=|-_{\theta}\rangle_3\langle-_{\theta}|$
($M^{(3)}_-=|-\rangle_3\langle-|$), where the measurement basis
$\{|+_\theta\rangle,|-_\theta\rangle\}$ is given in
Eq.~(\ref{basis}).}
\end{figure}

In practice, qubits 1, 2, and 3 will undergo independently
decoherence induced by local noises, and the canonical GHZ state
will be converted into a mixed state before we preforming the
entanglement localization procedure, as shown in Fig.~1 (b). We
first consider the amplitude noise \cite{Nielsen} in section A, and
then discuss other noise models, e.g., depolarizing model
\cite{Nielsen}, in section B.

\subsection{Entanglement localization under amplitude-damping decoherence}

Amplitude-damping decoherence is suited to many practical qubit
systems, including vacuum-single-photon qubit with photon loss,
atomic qubit with spontaneous decay, and superconducting qubit with
zero-temperature energy relaxation. The action of amplitude noise
can be described by two Krauss operators,
\begin{eqnarray}
  K_0=\left(
  \begin{array}{cc}
     1 & 0\\
     0 & \sqrt{\overline{d}}
  \end{array}
  \right),~~~~
  K_1=\left(
  \begin{array}{cc}
     0 & \sqrt{d}\\
     0 & 0
  \end{array}
  \right)
\end{eqnarray}
with $0\leqslant d\leqslant1$ and $\bar{d}=1-d$. $K_1$ describes the
transition of $|1\rangle$ to $|0\rangle$, while $K_0$ describes the
evolution of the system without such a transition. Note that $d=0$
denotes the noise-free case and $d=1$ means the interactional time
or strength between the system and environment tending to infinity.
Therefore, the decoherence strength $d$ is acquiesced in the range
$(0,1)$ in the following discussion.

After each qubit interacting with a local amplitude-damping
environment, the standard GHZ state in Eq.~(\ref{GHZ}) degenerates
to a mixed state
\begin{eqnarray}
 \rho^{(123)}&=&\sum\limits_{l,m,n=0}^1K_l\otimes K_m\otimes K_n|\psi\rangle^{(123)}\langle \psi|K_l^+\otimes K_m^+\otimes K_n^+\nonumber\\
  &=& \frac{1}{2}(1+d_1d_2d_3)|000\rangle\langle000|+\frac{1}{2}\bar{d}_1\bar{d}_2\bar{d}_3|111\rangle\langle111|\nonumber\\
  &&+\frac{1}{2}\sqrt{\bar{d}_1\bar{d}_2\bar{d}_3}|000\rangle\langle111|+\frac{1}{2}\sqrt{\bar{d}_1\bar{d}_2\bar{d}_3}|111\rangle\langle000|\nonumber\\
  &&+\frac{1}{2}d_1d_2\bar{d}_3|001\rangle\langle001|+\frac{1}{2}d_1\bar{d}_2d_3|010\rangle\langle010|\nonumber\\
  &&+\frac{1}{2}\bar{d}_1d_2d_3|100\rangle\langle100|+\frac{1}{2}d_1\bar{d}_2\bar{d}_3|011\rangle\langle011|\nonumber\\
  &&+\frac{1}{2}\bar{d}_1d_2\bar{d}_3|101\rangle\langle101|+\frac{1}{2}\bar{d}_1\bar{d}_2d_3|110\rangle\langle110|,
\end{eqnarray}
where $d_1$, $d_2$, and $d_3$ denote the decoherence strengths of
qubits 1, 2, and 3, respectively. For helping Alice and Bob to
establish a two-qubit entangled state with as much entanglement as
possible, Charlie needs to make a suitable local measurement on
qubit 3 and informs them of the outcome. We here only pay attention
to the von Neumann measurement. The general single-qubit projective
measurement basis can be described by
\begin{eqnarray}
\label{basis}
 |+_{\theta}\rangle&=&\cos\frac{\theta}{2}|0\rangle+\sin\frac{\theta}{2}e^{i\varphi}|1\rangle,\nonumber\\
 |-_{\theta}\rangle&=&\sin\frac{\theta}{2}e^{-i\varphi}|0\rangle-\cos\frac{\theta}{2}|1\rangle,
\end{eqnarray}
where $\theta\in[0,\pi]$ and $\varphi\in[0,2\pi]$. When
$\theta=\pi/2$ and $\varphi=0$, $|\pm_\theta\rangle$ reduce to
$|\pm\rangle$. The probability of getting the outcome
$|+_{\theta}\rangle_3$ is given by
\begin{eqnarray}
\label{P+}
 P_+(d,\theta)&=&\mathrm{Tr}\left[|+_\theta\rangle_3\langle+_\theta|\rho^{(123)}\right]\nonumber\\
   &=&\frac{1}{2}+\frac{d_3}{2}\cos\theta.
\end{eqnarray}
The occurrence of this event will lead to the fact that qubits 1 and
2 are projected in the state
\begin{eqnarray}
 \rho^{(12)}_+&=&\frac{1}{P_+}\mathrm{Tr}_3\left[|+\rangle_3\langle+|\rho^{(123)}\right]\nonumber\\
 &=&\frac{1}{P_+}\left(\gamma_+|00\rangle\langle00|+\kappa_+|01\rangle\langle01|+\tau_+|10\rangle\langle10|\right.\nonumber\\
 &&\left.+\eta_+|11\rangle\langle11|+\xi|00\rangle\langle11|+\xi^*|11\rangle\langle00|\right),
 \label{rhop}
\end{eqnarray}
where
\begin{eqnarray}
\gamma_+ &=& \frac{1}{2}(1+d_1d_2d_3)\cos^2\frac{\theta}{2}+\frac{1}{2}d_1d_2\bar{d}_3\sin^2\frac{\theta}{2},\nonumber\\
\kappa_+&=&\frac{1}{2}d_1\bar{d}_2d_3\cos^2\frac{\theta}{2}+\frac{1}{2}d_1\bar{d}_2\bar{d}_3\sin^2\frac{\theta}{2},\nonumber\\
\tau_+&=&\frac{1}{2}\bar{d}_1d_2d_3\cos^2\frac{\theta}{2}+\frac{1}{2}\bar{d}_1d_2\bar{d}_3\sin^2\frac{\theta}{2},\nonumber\\
\eta_+&=&\frac{1}{2}\bar{d}_1\bar{d}_2d_3\cos^2\frac{\theta}{2}+\frac{1}{2}\bar{d}_1\bar{d}_2\bar{d}_3\sin^2\frac{\theta}{2},\nonumber\\
\xi&=&\frac{1}{2}\sqrt{\bar{d}_1\bar{d}_2\bar{d}_3}\sin\frac{\theta}{2}\cos\frac{\theta}{2}e^{i\varphi}.
\end{eqnarray}
If the measurement outcome on qubit 3 is $|-_{\theta}\rangle_3$,
which happens with probability
\begin{eqnarray}
P_-(d,\theta)&=&\mathrm{Tr}\left[|-\rangle_3\langle-|\rho^{(123)}\right]=1-P_+(d,\theta)\nonumber\\
   &=&\frac{1}{2}-\frac{d_3}{2}\cos\theta,
\end{eqnarray}
qubits 1 and 2 will be projected in the state
\begin{eqnarray}
\rho^{(12)}_-&=&\frac{1}{P_-}\mathrm{Tr}_3\left[|-\rangle_3\langle-|\rho^{(123)}\right]\nonumber\\
 &=&\frac{1}{P_-}\left(\gamma_-|00\rangle\langle00|+\kappa_-|01\rangle\langle01|+\tau_-|10\rangle\langle10|\right.\nonumber\\
 &&\left.+\eta_-|11\rangle\langle11|-\xi|00\rangle\langle11|-\xi^*|11\rangle\langle00|\right),
 \label{rhom}
\end{eqnarray}
where
\begin{eqnarray}
\gamma_- &=& \frac{1}{2}(1+d_1d_2d_3)\sin^2\frac{\theta}{2}+\frac{1}{2}d_1d_2\bar{d}_3\cos^2\frac{\theta}{2},\nonumber\\
\kappa_-&=&\frac{1}{2}d_1\bar{d}_2d_3\sin^2\frac{\theta}{2}+\frac{1}{2}d_1\bar{d}_2\bar{d}_3\cos^2\frac{\theta}{2},\nonumber\\
\tau_-&=&\frac{1}{2}\bar{d}_1d_2d_3\sin^2\frac{\theta}{2}+\frac{1}{2}\bar{d}_1d_2\bar{d}_3\cos^2\frac{\theta}{2},\nonumber\\
\eta_-&=&\frac{1}{2}\bar{d}_1\bar{d}_2d_3\sin^2\frac{\theta}{2}+\frac{1}{2}\bar{d}_1\bar{d}_2\bar{d}_3\cos^2\frac{\theta}{2}.
\end{eqnarray}

Next, we use two measures, negativity \cite{58PRL883,65PRA032314}
and fully entangled fraction (FEF) \cite{54PRA3824,60PRA1888}, to
quantify the entanglement of $\rho_+^{(12)}$ and $\rho_-^{(12)}$,
respectively, and analyze their features. Negativity has been
considered as a dependable measure of entanglement for bipartite
entangled states \cite{58PRL883,65PRA032314}. FEF, which expresses
the purity of a bipartite mixed state, plays a central role in
quantum teleportation and entanglement distillation
\cite{54PRA3824,60PRA1888,76PRL722,78PRL574}, and may behave
differently from negativity as shown later.

\subsubsection{Negativity of the collapsed state of qubits 1 and 2}

Following Ref.~\cite{58PRL883}, we use the following definition of
negativity:
\begin{equation}
 N(\rho)=\max\left\{0,-2\lambda_{\mathrm{min}}(\rho)\right\},
\end{equation}
with $\lambda_{\mathrm{min}}$ the minimal eigenvalue of the partial
transpose of $\rho$ denoted as $\rho^T$. After straightforward
calculations we obtain the negativity of $\rho_+^{(12)}$ and
$\rho_-^{(12)}$ as
\begin{eqnarray}
\label{Npg}
 N_+(\rho_+)=\max\left\{0,-2\mu_+\right\},\\
 N_-(\rho_-)=\max\left\{0,-2\mu_-\right\},
 \label{Nmg}
\end{eqnarray}
where $\mu_+$ and $\mu_-$ are, respectively, the minimal eigenvalues
of $\rho_+$ and $\rho_-$, given by
\begin{eqnarray}
 \mu_+=\frac{1}{2P_+}\left(\kappa_++\tau_+-\sqrt{(\kappa_+-\tau_+)^2+4|\xi|^2}\right),\\
 \mu_-=\frac{1}{2P_-}\left(\kappa_-+\tau_--\sqrt{(\kappa_--\tau_-)^2+4|\xi|^2}\right).
\end{eqnarray}
For clarity, we give a detailed analysis on $N_+$ and $N_-$ for the
case $d_1=d_2=d_3=d$ (which is not a necessary assumption but only
simplifies the degree of algebraic complexity). In this case, $\mu_+$ and
$\mu_-$ reduce, respectively, to
\begin{eqnarray}
\mu^s_+=\frac{\bar{d}}{2P_+}\left(d^2\cos^2\frac{\theta}{2}+d\bar{d}\sin^2\frac{\theta}{2}-\sqrt{\bar{d}~}\sin\frac{\theta}{2}\cos\frac{\theta}{2}\right),\\
\mu^s_-=\frac{\bar{d}}{2P_-}\left(d^2\sin^2\frac{\theta}{2}+d\bar{d}\cos^2\frac{\theta}{2}-\sqrt{\bar{d}~}\sin\frac{\theta}{2}\cos\frac{\theta}{2}\right).
\end{eqnarray}
The clear dependence of $N_+$ and $N_-$ on $d$ and $\theta$ is
plotted in Fig.~2.
\begin{figure}
\includegraphics[width=8cm,height=6cm]{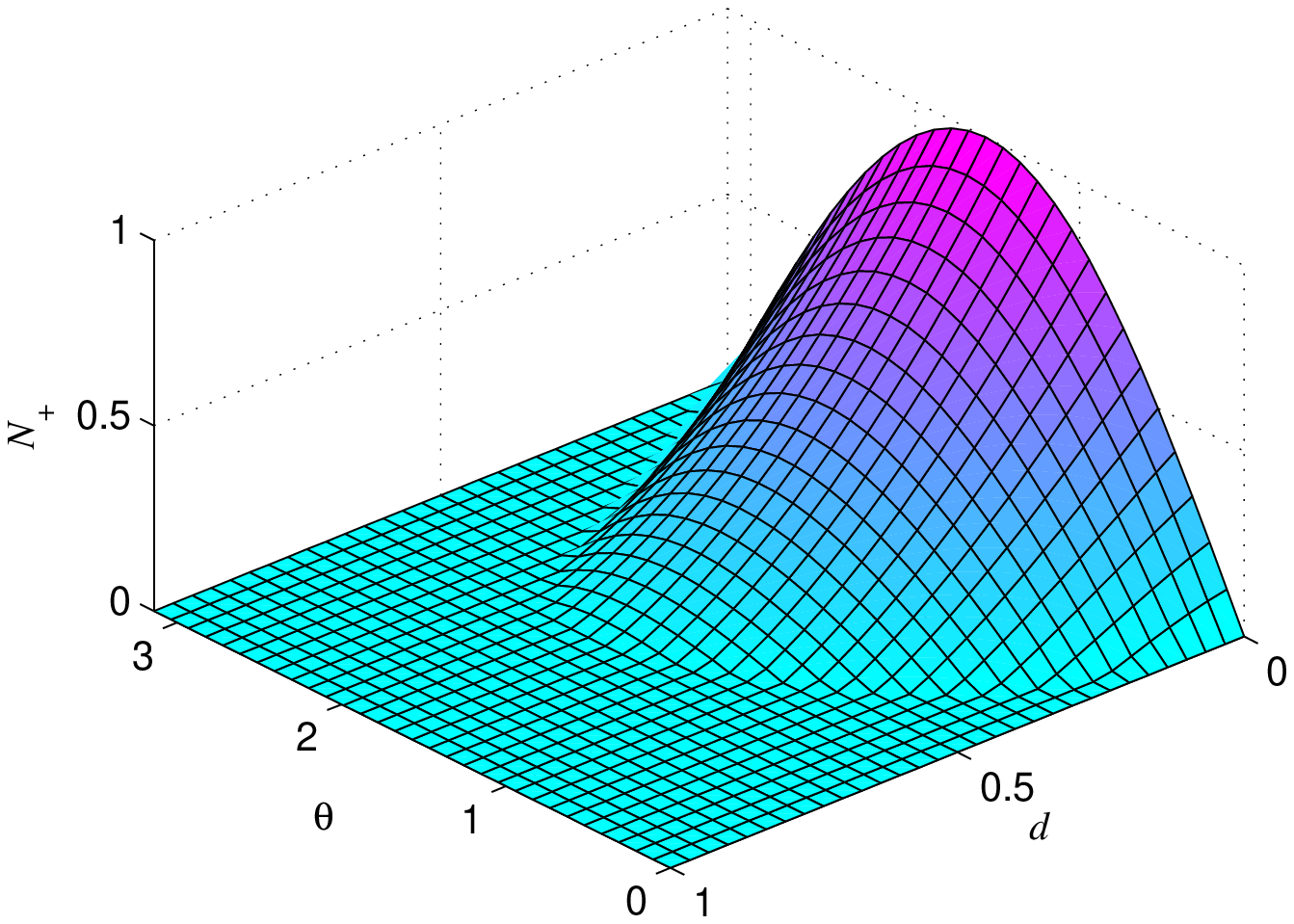}\\
\includegraphics[width=8cm,height=6cm]{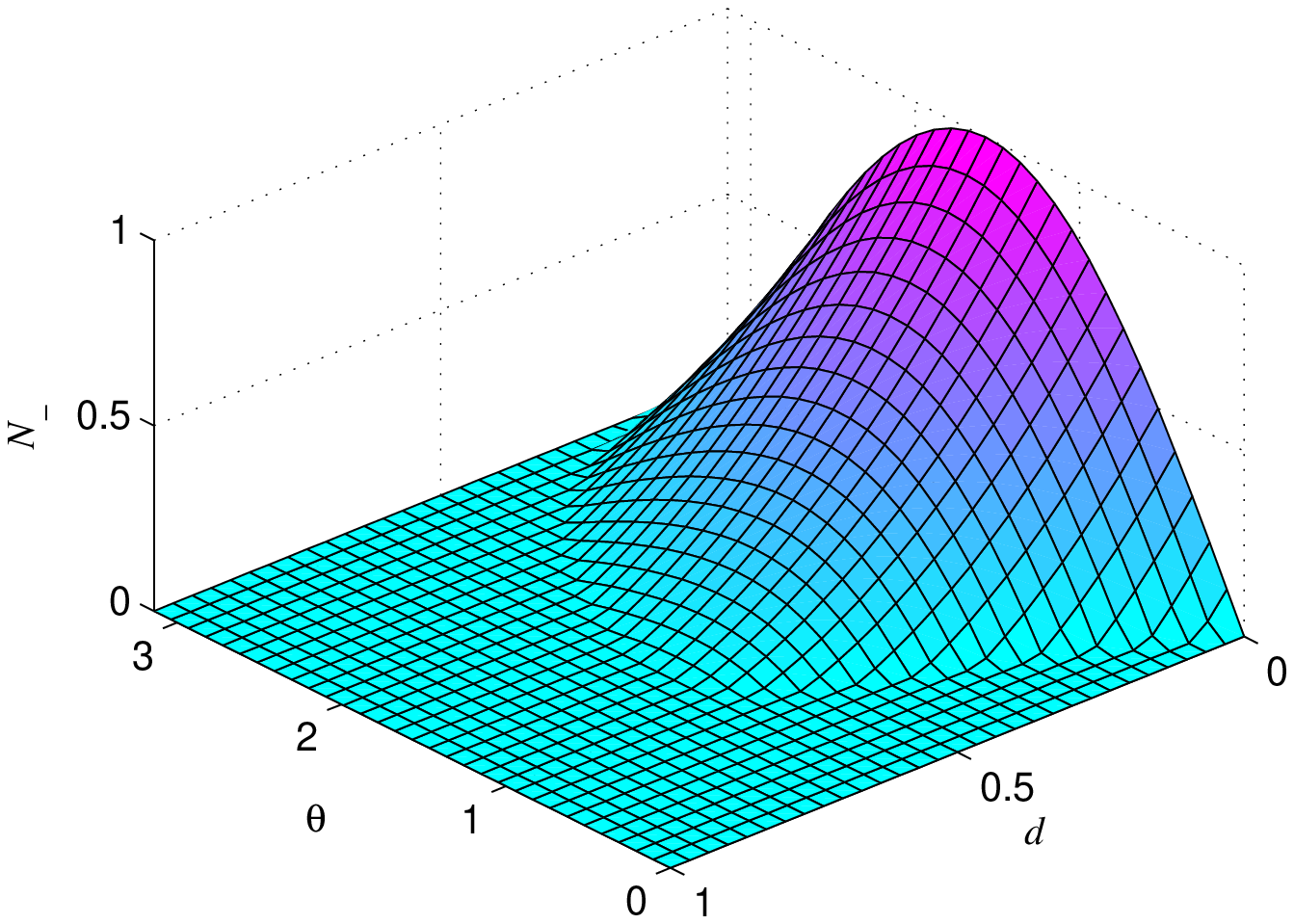}
\caption{(Color online) The function graph of $N_+$ and $N_-$ with
respect to $d$ and $\theta$.}
\end{figure}

It can be seen from Fig.~2 that when $d$ increases to a threshold,
being away from one, for a given $\theta$, both $N_+$ and $N_-$
decrease to zero. This indicates that the entanglement vanishes in a
finite time, which is referred to as entanglement sudden death
\cite{316S579,93PRL140404,99PRL180504,99PRL160502,76PRA062322}. More
interesting and important information that can be obtained from
Fig.~2 is as follows. If $d=0$ (corresponding to the absence of
noise), both of $N_+$ and $N_-$ attain their maximal values at
$\theta=\pi/2$, meaning that $\{|\pm\rangle_3\}$ is the optimal
measurement basis. This result is in accordance with the discussion
before. For $d>0$, however, both $N_+$ and $N_-$ are asymmetric with
respect to $\theta=\pi/2$ in the region that $d$ is less than the
threshold defined above. This feature implies that $N_+$ and $N_-$
reach their maximums at the points that deviate from $\theta=\pi/2$,
respectively. Such phenomena can be observed clearly in Fig.~3 which
gives the bivariate functions $\Delta
N_+(d,\theta)=N_+(d,\theta)-N_+(d,\theta=\pi/2)$ and $\Delta
N_-(d,\theta)=N_-(d,\theta)-N_-(d,\theta=\pi/2)$ with independent
variables $\theta$ and $d$. We can see that there exist different
regimes of $\theta$ in which $\Delta N_+$ and $\Delta N_-$ are
larger than zero, respectively; that is, $N_+(d,\theta\neq\pi/2)$
and $N_-(d,\theta\neq\pi/2)$ are indeed larger than $N_+(d,\pi/2)$
and $N_-(d,\pi/2)$, respectively. These results indicate that
Charlie can enhance probabilistically the entanglement distributed
between Alice and Bob by selecting an appropriate measurement basis
$\{|+_{\theta\neq\pi/2}\rangle,|-_{\theta\neq\pi/2}\rangle\}$
instead of $\{|+\rangle,|-\rangle\}$.

\begin{figure}
\includegraphics[width=8cm,height=6cm]{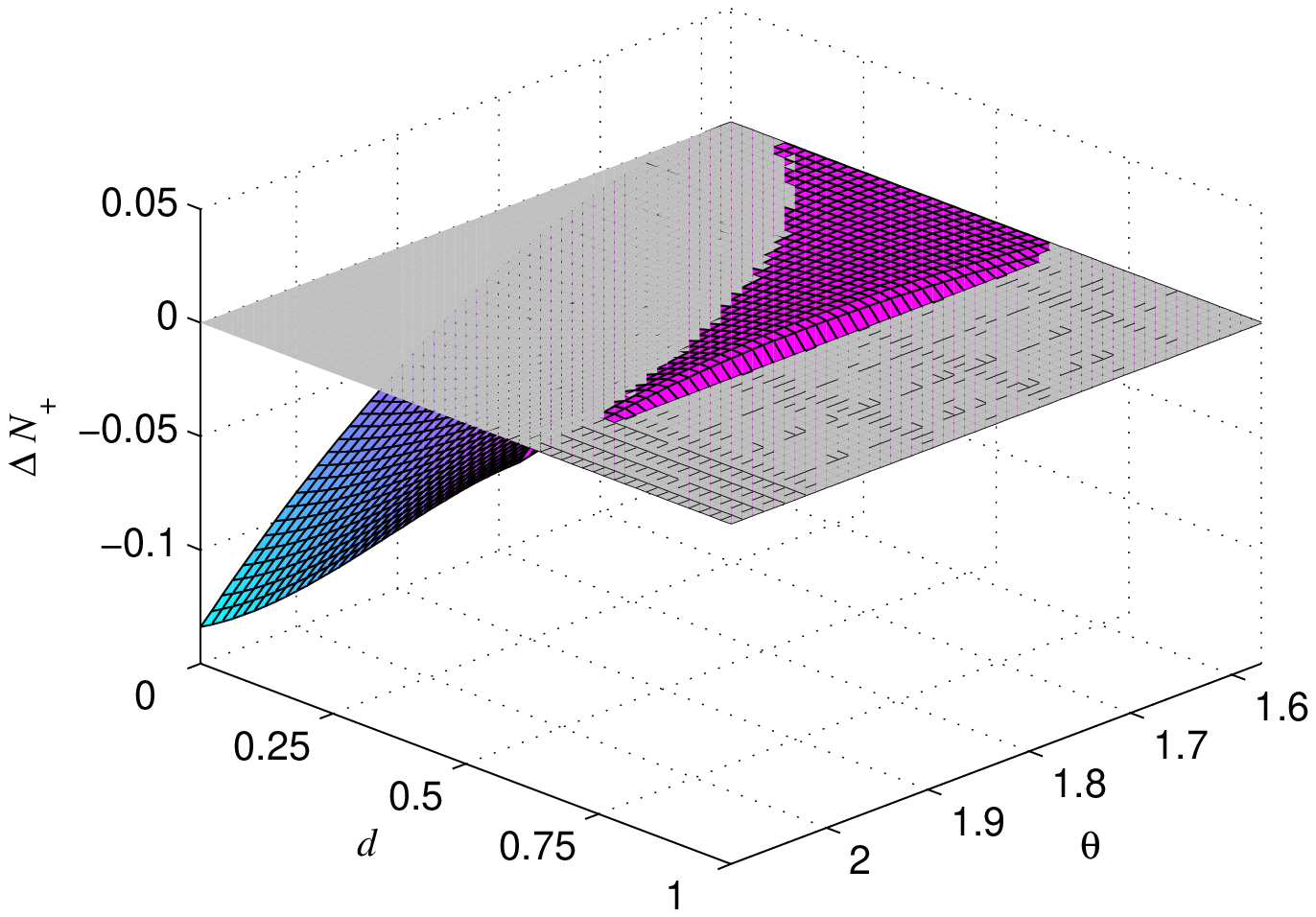}\\
\includegraphics[width=8cm,height=6cm]{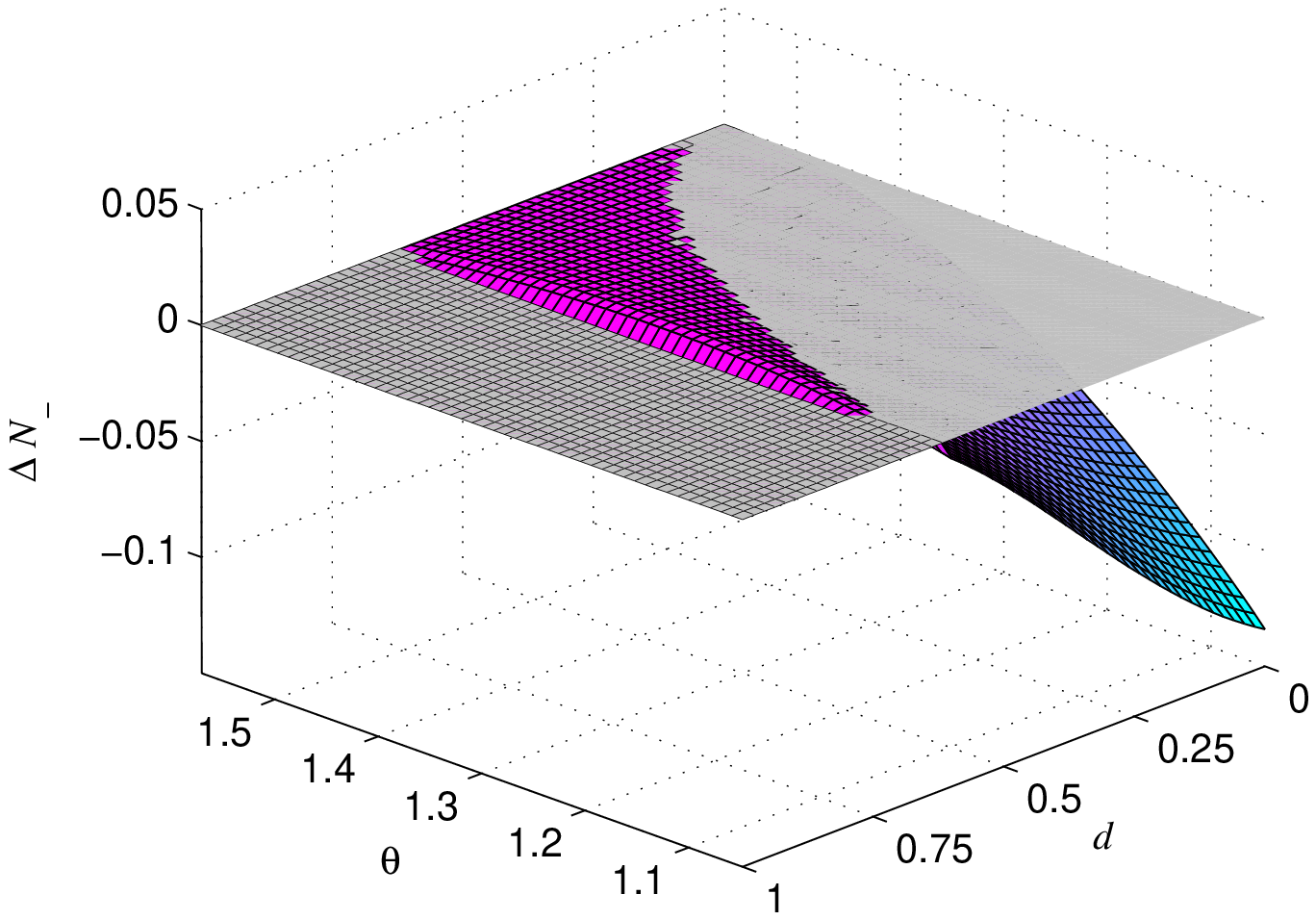}
\caption{(Color online) The dependence of $\Delta N_+$ and $\Delta
N_-$ on $d$ and $\theta$, where $\theta$ ranges from $\pi/2$ to
$2\pi/3$ in the upper graph and from $\pi/3$ to $\pi/2$ in the lower
graph.}
\end{figure}

The average amount of entanglement between qubits 1 and 2 for two
possible measurement outcomes $|+_\theta\rangle$ and
$|-_\theta\rangle$ is given by
\begin{equation}
 N_{\mathrm{ave}}(d,\theta)=P_+N_++P_-N_-.
\end{equation}
It can be easily verified that when $d$ is smaller than a threshold,
the maximal value of $N_{\mathrm{ave}}(d,\theta)$ is
$N_{\mathrm{ave}}(d,\pi/2)$ for a given $d$. When $d$ goes beyond
the threshold, however, $N_{\mathrm{ave}}(d,\theta)$ can attain its
maximum at two different values of $\theta$, situating symmetrically
on the two sides of $\theta=\pi/2$, provided that
$N_{\mathrm{ave}}(d,\theta)$ is not always equal to zero, as shown
in Fig.~4; this fact means that $\{|\pm\rangle\}$ is no longer the
optimal measurement basis of qubit 3. Figure 4 also indicates that
the existing time of the entanglement of the state $\rho_+$ or
$\rho_-$ can be protracted by taking an appropriate measurement
basis $\{|+_{\theta\neq\pi/2}\rangle,|-_{\theta\neq\pi/2}\rangle\}$
instead of $\{|+\rangle,|-\rangle\}$.

\begin{figure}
\includegraphics[width=8cm,height=6cm]{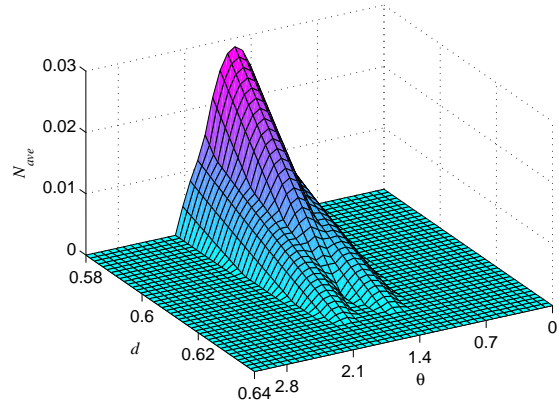}
\caption{(Color online) $N_{\mathrm{ave}}$ versus $d$ and $\theta$.
The figure only plots the region of $0.58\leqslant d\leqslant
0.64$.}
\end{figure}

In the discussion above, we supposed that everyone of three qubits
suffer decoherence. The obtained result is naturally applicable to
special cases. That is, when only one or two qubits sustain
decoherence, the optimal measurement basis of qubit 3 is also not
$\{|+\rangle,|-\rangle\}$. Let us take an example of $d_3=0$ and
$d_1=d_2=d$. Then $N_+$ and $N_-$ in Eqs.~(\ref{Npg}) and
(\ref{Nmg}) reduce, respectively, to
\begin{eqnarray}
\label{Npe}
 N_+=\max\left\{0,~2\bar{d}\sin\frac{\theta}{2}\left(\cos\frac{\theta}{2}-d\sin\frac{\theta}{2}\right)\right\},\\
 N_-=\max\left\{0,~2\bar{d}\cos\frac{\theta}{2}\left(\sin\frac{\theta}{2}-d\cos\frac{\theta}{2}\right)\right\},
 \label{Nme}
\end{eqnarray}
Evidently, the points of maximum of both $N_+$ and $N_-$ are not at
$\theta=\pi/2$. That is to say, the best measurement basis of qubit
3 is not $\{|+\rangle,|-\rangle\}$ in the aforementioned
entanglement localization protocol.

\subsubsection{FEF of the collapsed state of qubits 1 and 2}

FEF of a state $\rho$ is defined as the maximum overlap of $\rho$
with a maximally entangled state \cite{54PRA3824,60PRA1888}, that
is,
\begin{equation}
F(\rho)=\max\limits_{|\phi\rangle}\langle\phi|\rho|\phi\rangle,
\end{equation}
where the maximization is taken over all maximally entangled states
$|\phi\rangle$. For two-qubit systems $F(\rho)$ can be analytically
expressed as \cite{62PRA012311}
\begin{equation}
F(\rho)=\frac{1}{4}\left\{1+\mu_1+\mu_2-\mathrm{sgn}[\det(\tilde{R})]\mu_3\right\},
\end{equation}
where $\{\mu_i\}$ are the decreasingly ordered singular values of
the $3\times 3$ real matrix
$\tilde{R}=\left[\mathrm{tr}(\rho\sigma_i\otimes\sigma_j)\right]_{3\times3}$
with $\{\sigma_i,i=1,2,3\}$ the Pauli matrices and
$\mathrm{sgn}[\det(\tilde{R})]$ is the sign of the determinant of
$\tilde{R}$.

The FEF of the states $\rho_+$ and $\rho_-$ in Eq.~(\ref{rhop}) and
Eq.~(\ref{rhom}) can be calculated to be
\begin{eqnarray}
\label{Fpg}
 F_+(\rho_+)&=&\frac{1}{4}+\frac{\left(4|\xi|+\gamma_++\eta_+-\kappa_+-\tau_+\right)}{4P_+},\\
 F_-(\rho_-)&=&\frac{1}{4}+\frac{\left(4|\xi|+\gamma_-+\eta_--\kappa_--\tau_-\right)}{4P_-}.
 \label{Fmg}
\end{eqnarray}
As before, we still discuss the case $d_1=d_2=d_3=d$. Then
$F_+(\rho_+)$ and $F_-(\rho_-)$ reduce, respectively, to
\begin{eqnarray}
 F_+(d,\theta)&=&\frac{1}{2}-\mu^s_+, \label{Fp}\\
 F_-(d,\theta)&=&\frac{1}{2}-\mu^s_-.
 \label{Fm}
\end{eqnarray}
Then the FEF $F_+$ and $F_-$ have the similar behaviors to the
negativity $N_+$ and $N_-$, respectively. That is, $F_+$ and $F_-$
reach their maximal values at $\theta\neq\pi/2$. As a matter of
fact, $F_+$ ($F_-$) and $N_+$ ($N_-$) have the same extremal point,
and there exist the same scale of $d$ in which $F_+(d,\theta)$
[$F_-(d,\theta)$] and $N_+(d,\theta)$ [$N_-(d,\theta)$] are larger
than $F_+(d,\theta=\pi/2)$ [$F_-(d,\theta=\pi/2)$] and
$N_+(d,\theta=\pi/2)$ [$N_-(d,\theta=\pi/2)$], respectively. Thus
Charlie can also increase the FEF of the state shared by Alice and
Bob by adopting a suitable measurement basis
$\{|+_{\theta\neq\pi/2}\rangle,|-_{\theta\neq\pi/2}\rangle\}$
instead of $\{|+\rangle,|-\rangle\}$.

The mean value of $F_+$ and $F_-$ can be calculated as
\begin{eqnarray}
 F_{\mathrm{ave}}&=&P_+F_+ +P_-F_- \nonumber\\
 &=&\frac{3}{8}+\sqrt{\bar{d}_1\bar{d}_2\bar{d}_3}\sin\frac{\theta}{2}\cos\frac{\theta}{2}+\frac{1}{8}(2d_1-1)(2d_2-1).\nonumber\\
\end{eqnarray}
For $d_1=d_2=d_3=d$, $F_{\mathrm{ave}}$ reduces to
\begin{eqnarray}
 F_{\mathrm{ave}}(d,\theta)=\frac{3}{8}+\bar{d}\sqrt{\bar{d}~}\sin\frac{\theta}{2}\cos\frac{\theta}{2}+\frac{1}{8}(2d-1)^2.
 \label{ave}
\end{eqnarray}
Obviously, the maximal value of $F_{\mathrm{ave}}(d,\theta)$ is
$F_{\mathrm{ave}}^{\mathrm{max}}(d)=F_{\mathrm{ave}}(d,\theta=\pi/2)$
which is independent of the parameter $\theta$. This result
indicates that $F_{\mathrm{ave}}$ has different behavior to
$N_{\mathrm{ave}}(d,\theta)$ which reaches the maximal value at
$\theta\neq \pi/2$ when $d$ oversteps a critical value (see Fig.~4).

In view of practice, however, what we are interested in is to
maximize $F_+$ or $F_-$, due to the fact that the larger the FEF is,
the higher teleportation fidelity and entanglement purification
efficiency can be achieved
\cite{54PRA3824,60PRA1888,76PRL722,78PRL574}. Moreover, we notice
that if and only if the FEF of a two-qubit state $\rho$ is larger
than 1/2, quantum teleportation can exhibit its superiority over
state estimation based on classical strategies and entanglement
purification can be carried out effectively using the resource state
$\rho$ \cite{54PRA3824,60PRA1888,76PRL722,78PRL574}. We observe that
$F_{\mathrm{ave}}(d,\theta)\leqslant1/2$ does not mean
$F_+(d,\theta)$ and $F_-(d,\theta)$ are simultaneously less than
$1/2$. In deed, when $d\geqslant\left(\sqrt{5}-1\right)/2$,
$F_{\mathrm{ave}}^{\mathrm{max}}\leqslant1/2$ [obtained from
Eq.~(\ref{ave})], indicating that the resource state is useless for
quantum teleportation and entanglement distillation, while
$F_+(d,\theta>\pi/2)$ or $F_-(d,\theta<\pi/2)$ can overtop $1/2$ as
displayed in Fig.~5. Thus we could safely conclude that when we take
the measurement strategy that maximizes $F_{\mathrm{ave}}$, both
$\rho_+$ and $\rho_-$ may be useless for quantum teleportation and
entanglement distillation; in contrast, if we select an appropriate
measurement basis $\{|\pm_{\theta\neq\pi/2}\rangle\}$ rather than
$\{|\pm\rangle\}$ such that
$F_{\mathrm{ave}}<F_{\mathrm{ave}}^{\mathrm{max}}$, Alice and Bob
can implement effective teleportation and entanglement distillation
with a nonzero probability. In other words, $\{|\pm\rangle_3\}$ is
not the best measurement basis for optimizing the robustness of the
entangled state of qubits 1 and 2.

It has been mentioned before that maximizing the average amount of
entanglement between two particles of a multiparticle state by
performing local measurements on the other particles is defined as
localizable-entanglement \cite{92PRL027901,71PRA042306}. The
conclusions presented above imply that localizable-entanglement is
not suitable to be described by the entanglement measure of FEF from
the practical point of view.

\begin{figure}
\includegraphics[width=8cm,height=6cm]{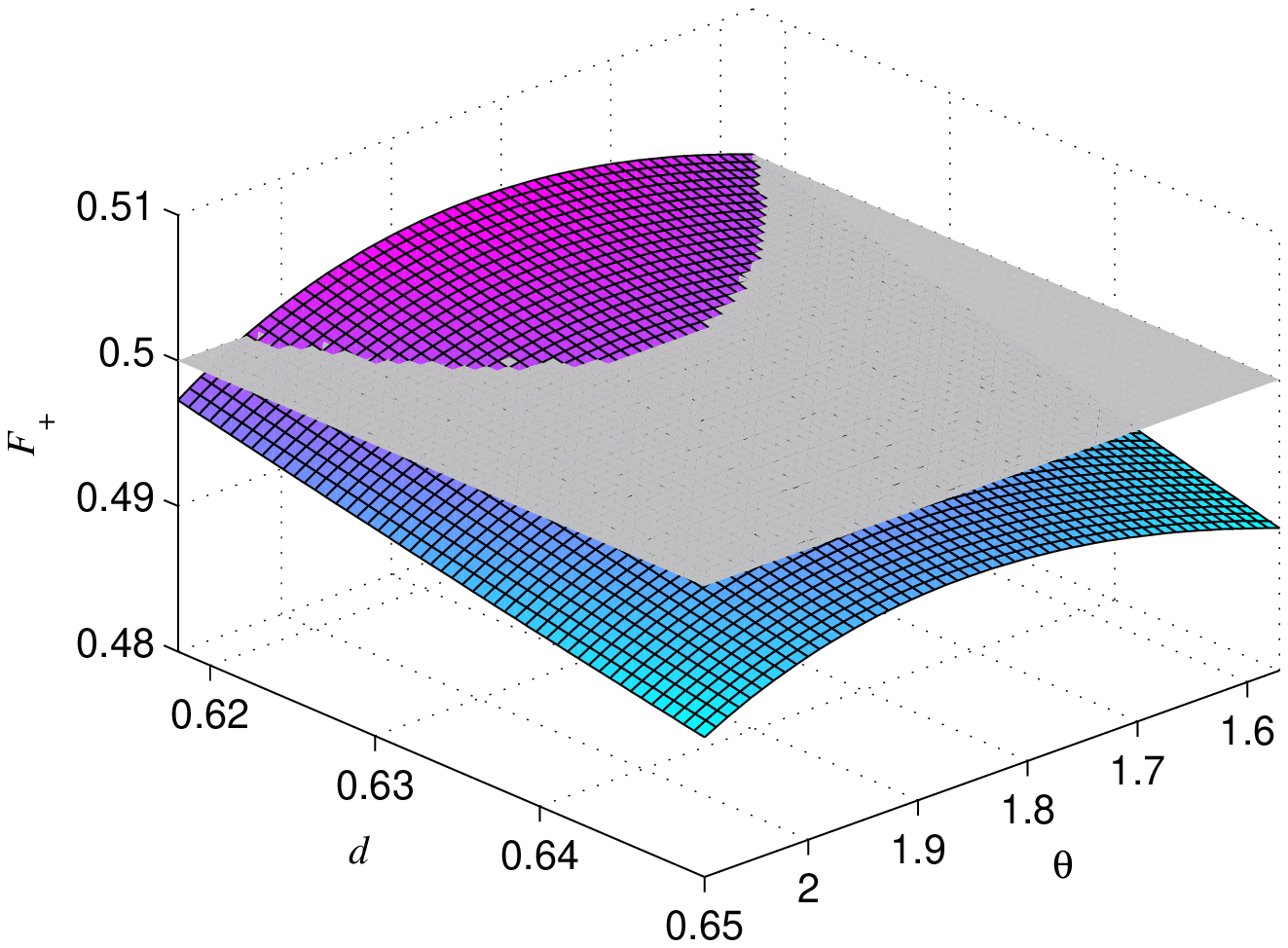}\\
\includegraphics[width=8cm,height=6cm]{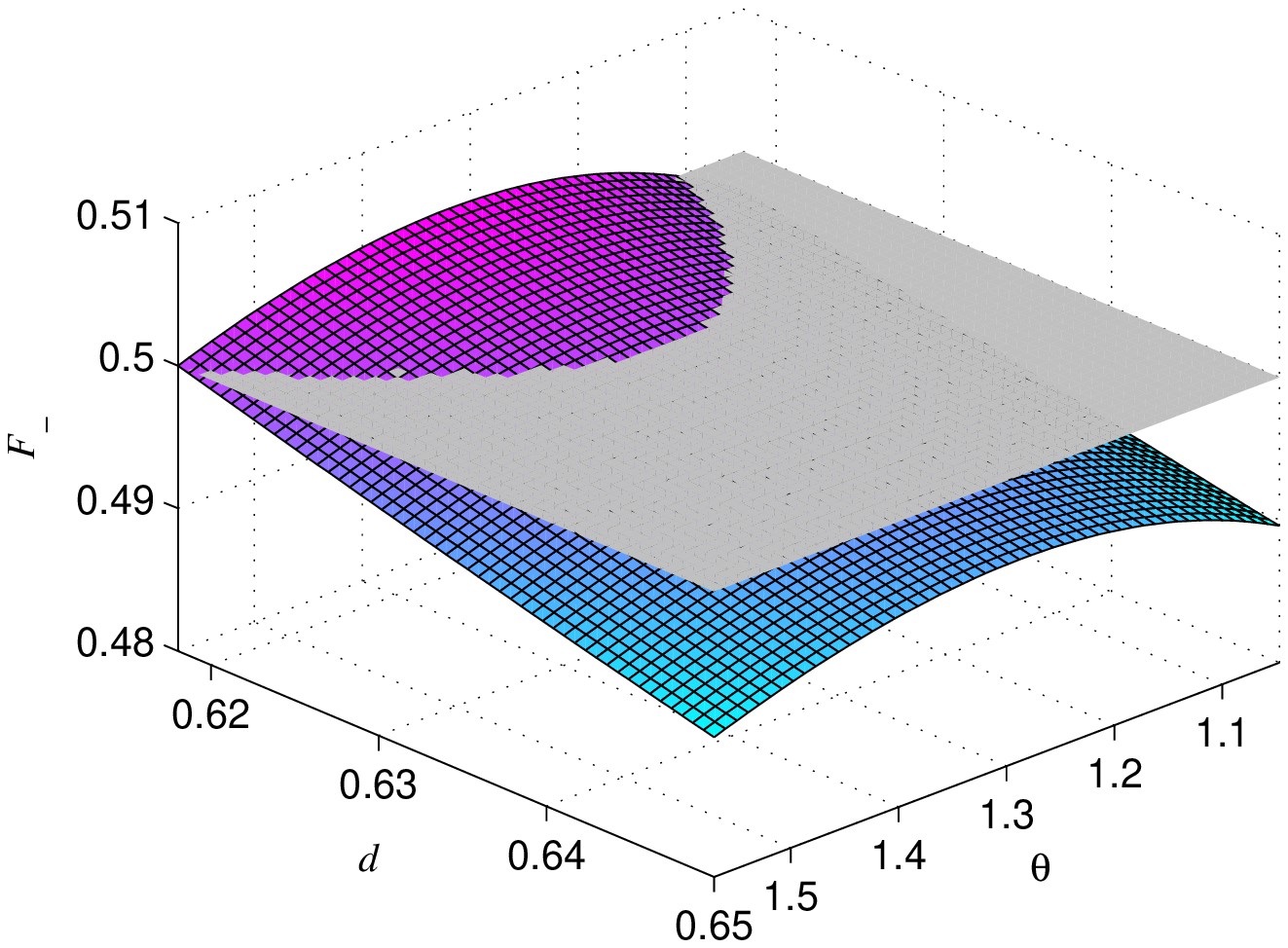}
\caption{(Color online) $F_+$ and $F_-$ versus $d$ and $\theta$,
where $d$ is in the range $[(\sqrt{5}-1)/2,0.65]$ and $\theta$ is in
the range $[\pi/2,2\pi/3]$ for the upper diagram and $[\pi/3,\pi/2]$
for the lower diagram.}
\end{figure}

Although FEF may be not monotonic in the regime of small values
under trace-preserving local operations and classical communication
(TPLOCC) for mixed states
\cite{62PRA012311,65PRA022302,90PRL097901,86PRA020304}, the
aforesaid conclusions are reliable as explained below. The
expressions of FEF in Eq.~(\ref{Fp}) and Eq.~(\ref{Fm}) can be
rewritten as
\begin{eqnarray}
 F_+(d,\theta)=\left\{
\begin{array}{ll}
 \frac{1}{2}(1-2\mu^s_+)&~~~\mathrm{for}~~N_+=0~~(\mu^s_+\geqslant0),\\\\
 \frac{1}{2}(1+N_+)&~~~\mathrm{for}~~N_+>0~~(\mu^s_+<0),
\end{array}
\right.\\
 F_-(d,\theta)=\left\{
\begin{array}{ll}
 \frac{1}{2}(1-2\mu^s_-)&~~~\mathrm{for}~~N_-=0~~(\mu^s_-\geqslant0),\\\\
 \frac{1}{2}(1+N_-)&~~~\mathrm{for}~~N_->0~~(\mu^s_-<0).
\end{array}
\right.
 \label{Fpmr}
\end{eqnarray}
It is well known that TPLOCC cannot create entanglement and thus
cannot increase $N_+$ or $N_-$. Then, TPLOCC cannot increase $F_+$
or $F_-$ in the present case because of the relationships given in
Eqs.~(29) and (30), which is in accordance with the result of
Ref.~\cite{66PRA022307}. In fact, both states $\rho_+$ and $\rho_-$
do not belong to the class of states presented in
Refs.~\cite{62PRA012311,65PRA022302} whose FEF may be slightly
raised by TPLOCC operations. The same argument could be obtained for
the results in the following context. Thus we will not consider the
local manipulations on qubits 1 and 2 and the classical
communications between Alice and Bob themselves later.

\subsection{Entanglement localization under depolarizing decoherence}

In the former subsection, we have shown that the optimal strategy
for extracting a two-qubit entangled state from a three-qubit GHZ
state via local measurements in the amplitude-damping case is
different from that in the noise-free case. Particularly, in the
ideal case, the best measurement basis of qubit 3 for reducing the
three-qubit GHZ state $|\psi\rangle^{(123)}$ to a two-qubit
entangled state $\rho^{(12)}$ is $\{|+\rangle_3,|-\rangle_3\}$;
while considering the amplitude-damping decoherence of part or all
of these qubits, the best measurement basis of qubit 3 is no longer
$\{|+\rangle_3,|-\rangle_3\}$. This phenomenon does not necessarily
occur under other noise models. We here take an example of
depolarizing model.

The single-qubit depolarizing channel is described as
\begin{equation}
\label{depolarizing}
\mathcal{E}(\rho)=\sum\limits_{i=0}^3p_i\sigma_i\rho \sigma_i
\end{equation}
where $\rho$ is the input state of the qubit, $p_0=1-d$ and
$p_i=d/3$ ($i = 1,2,3$) with $d$ being the degree of decoherence
($0\leq d\leq 1$), $\sigma_0$ is the identity operator, and
$\{\sigma_i\}$ ($i = 1,2,3$) are the Pauli operators $\sigma_x$,
$\sigma_y$, $\sigma_z$, respectively.

For the initial three-qubit GHZ state $|\psi\rangle^{(123)}$ in
Eq.~(\ref{GHZ}), the depolarizing operation on each qubit will
result in it becoming
\begin{equation}
\rho\prime^{(123)}=\sum\limits_{i,j,k=0}^3p_ip_jp_k\sigma_i\otimes\sigma_j\otimes\sigma_k|\psi\rangle\langle\psi|\sigma_i\otimes\sigma_j\otimes\sigma_k.
\end{equation}
Without loss of generality, we consider that the degree of
decoherence of every qubit is not zero. For simplicity, we suppose
the qubits have the same degree of decoherence $d$. After the
aforementioned entanglement localization process, the negativity of
the final state of qubits 1 and 2 is
\begin{equation}
 \mathcal{N}=\max\left\{0,-2\lambda\right\},
\end{equation}
where $\lambda=2(3-2d)d/9-|3-4d|^3\sin\theta/54$ for both the
measurement outcomes $|+_{\theta}\rangle_3$ and
$|-_{\theta}\rangle_3$ of qubit 3. In order to guarantee
$\mathcal{N}>0$, the condition
\begin{equation}
\label{condition}
  \sin\theta>\frac{12(3-2d)d}{|3-4d|^3}
\end{equation}
should be satisfied. Then the point of maximum of $\mathcal{N}$ is
at $\theta=\pi/2$ for any $d$. Moreover, the condition of
Eq.~(\ref{condition}) in the case $\theta=\pi/2$ can be satisfied
more easily than in the case $\theta\neq\pi/2$. Thus, the optimal
measurement basis of qubit 3 is
$\{|+_{\theta=\pi/2}\rangle=|+\rangle,|-_{\theta=\pi/2}\rangle=|-\rangle\}$.
Similarly, using the entanglement measure of FEF, the same
conclusion can be obtained. In a word, the optimal strategy for
reducing a three-qubit GHZ state to a two-qubit entangled state via
local measurements in the depolarizing case is the same as that in
the noise-free case.

The results above indicate that depolarizing and amplitude-damping
noises have different effects on entanglement localization. It tells
us that in different environments, we should take different
strategies for optimizing the entanglement localization schemes.

\section{Bipartite entanglement distribution assisted by three-particle entangled states}

Inspired by the afore-cited phenomena in section II, we find that
multiparticle entangled states could help to improve the quality of
entanglement distribution between two distant parties in noisy
environments, as demonstrated in this section.

A routine way of bipartite entanglement distributing between two
distant parties, Alice and Bob, is to generate a two-qubit entangled
state, e.g., a Bell state, in a server, say Charlie, and then
physically send the two qubits to the labs of Alice and Bob,
respectively. We here propose another way that first preparing a
three-qubit entangled state, GHZ state, in Charlie's site and then
send any two qubits, e.g., qubits 1 and 2, to Alice and Bob, one
person one qubit, followed by the entanglement localization
procedure introduced in the former section. In the noise-free case,
the two methods will achieve the same result in terms of the shared
entanglement between Alice and Bob. However, when considering the
unavoidable effect of noises on the systems during their
transmission, the latter scheme could boost probabilistically the
amount of entanglement of the two-qubit state shared by Alice and
Bob, as shown below. For clarity, the first method will be called
direct distribution scheme, DDS for short, and the second one will
be referred to as ancilla-assisted distribution scheme abbreviated
to ADS. The schematic diagrams of both DDS and ADS are sketched in
Fig.~6. The detailed descriptions on the DDS and ADS are given in
sections A and B, respectively.

\begin{figure}
\includegraphics[width=7cm,height=7cm]{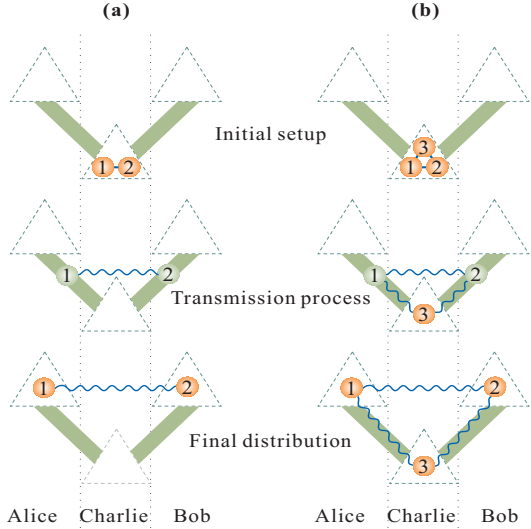} 
\caption{(Color online) Two schemes for distributing bipartite
entanglement. \textbf{(a)} Direct distribution scheme (DDS).
\textbf{(b)} Ancilla-assisted distribution scheme (ADS), where the
process of entanglement localization is not shown in the diagram and
the detailed description on it is given in the context (see also
Fig.~1). The green bars denote the quantum channels with which
Charlie sends particles 1 and 2 to Alice and Bob, respectively.
Linking the qubits by beelines denote that these qubits are in a
maximally entangled pure state, while linking the qubits by wave
lines denote these qubits being in a mixed state.}
\end{figure}

\subsection{DDS for distributing bipartite entanglement via noisy quantum channels}

In order to display the advantages of ADS later, we first
recapitulate the results of DDS for providing a sharp contrast.
Suppose that qubits 1 and 2 are initially prepared in a Bell state
\begin{equation}
 |\phi\rangle_{12}=\frac{1}{\sqrt{2}}(|00\rangle+|11\rangle)_{12}.
 \label{Bell}
\end{equation}
After the two qubits independently interacting with their
environments via amplitude-damping channels, the Bell state evolves
into a mixed state
\begin{eqnarray}
 \varrho_{12}&=&\sum\limits_{m,n=0}^1K_m\otimes K_n|\phi\rangle_{12}\langle \phi|K_m^+\otimes K_n^+\nonumber\\
  &=&\frac{1}{2}(1+d_1d_2)|00\rangle\langle00\rangle+\frac{1}{2}\bar{d}_1\bar{d}_2|11\rangle\langle11|\nonumber\\
  &&+\frac{1}{2}\sqrt{\bar{d}_1\bar{d}_2}|00\rangle\langle11|+\frac{1}{2}\sqrt{\bar{d}_1\bar{d}_2}|11\rangle\langle00|\nonumber\\
  &&+\frac{1}{2}d_1\bar{d}_2|01\rangle\langle01|+\frac{1}{2}\bar{d}_1d_2|10\rangle\langle10|.
\end{eqnarray}
The negativity and FEF of $\varrho$ can be calculated, respectively,
to be
\begin{eqnarray}
 N(\varrho)&=&\max\left\{0,-2\lambda_{\mathrm{min}}(\varrho)\right\},\\
 \lambda_{\mathrm{min}}&=&\frac{1}{4}\left(d_1\bar{d}_2+\bar{d}_1d_2\right)-\frac{1}{4}\sqrt{(d_1-d_2)^2+4\bar{d}_1\bar{d}_2} ,\nonumber\\
 F(\varrho)&=&\frac{1}{4}\left(2+2\sqrt{\bar{d}_1\bar{d}_2}+2d_1d_2-d_1-d_2\right).
 \label{Fsigma}
\end{eqnarray}
We assume $d_1=d_2=d$, that is, the decoherence strengths of both
qubits are the same. This is not a necessary assumption but only
simplifies the degree of algebraic complexity, which makes no
difference to the final conclusion. Then $N(\varrho)$ and
$F(\varrho)$ reduce to
\begin{eqnarray}
\label{Nvarrho}
N'(\varrho)&=&(1-d)^2,\\
F'(\varrho)&=&\frac{1}{2}\left(1+N'\right).
 \label{Fvarrho}
\end{eqnarray}

\subsection{ADS for distributing bipartite entanglement via noisy quantum channels}
Some results in Sec.~II can be transplanted to this section for
simplifying the discussion on the ADS of bipartite entanglement
distribution. It is observed from Sec.~II that the negativity and
FEF of the states $\rho_+$ and $\rho_-$ are symmetric about
$\theta=\pi/2$. Thus we here only discuss the entanglement
properties of $\rho_+$, and the counterparts for $\rho_-$ can be
directly obtained using the symmetry.

Following Eq.~(\ref{Npg}) and Eq.~(\ref{Fpg}), when $d_1=d_2=d$,
$N_+(\rho_+)$ and $F_+(\rho_+)$ reduce to
\begin{eqnarray}
\label{N'}
N'_+(\rho_+)&=&\max\left\{0,\frac{2}{P_+}\left(|\xi'|-\kappa'_+\right)\right\},\\
F'_+(\rho_+) &=&\left\{
\begin{array}{ll}
  \frac{1}{2}+\frac{1}{P_+}\left(\kappa'_+-|\xi'|\right)&~~\mathrm{for}~N_+=0,\\\\
 \frac{1}{2}(1+N'_+)&~~\mathrm{for}~N_+>0,
\end{array}
\right.\label{F'}
\end{eqnarray}
where
\begin{eqnarray}
 \kappa'_+&=&\frac{d\bar{d}}{2}\left(d_3\cos^2\frac{\theta}{2}+\bar{d}_3\sin^2\frac{\theta}{2}\right),\nonumber\\
 |\xi'|&=&\frac{\bar{d}}{2}\sqrt{\bar{d}_3}\sin\frac{\theta}{2}\cos\frac{\theta}{2}.
\end{eqnarray}

We now make a comparison between the aforementioned two strategies,
DDS and ADS, by analyzing the differences of the negativity and FEF
of the state $\rho_+$ with that of the state $\varrho$, which are
given by
\begin{eqnarray}
 \delta N&=&N'_+(\rho_+)-N'(\varrho),\\
 \delta F&=&F'_+(\rho_+)-F'(\varrho).
\end{eqnarray}
What we are interested in is whether $\delta N$ and $\delta F$ could
be larger than zero. This expectation is possible if and only if
$N'_+(\rho_+)>0$ and $F'_+(\rho_+)>1/2$. According to
Eqs.~(\ref{Nvarrho})-(\ref{F'}), it can be acquired that $\delta N$
and $\delta F$ have the same behavior in the regime of
$N'_+(\rho_+)>0$ and $F'_+(\rho_+)>1/2$. Thus we only need to
analyze the characteristics of $\delta N$, with which the features
of $\delta F$ can also be derived straightforwardly.

To exhibit ADS's superiority clearly, we first assume $d_3=0$,
meaning that qubit 3 is well isolated from the noisy environment in
Charlie's lab. In this case, the dependence of $\delta N$ on $d$ and
$\theta$ is given in Fig.~7 with $0\leqslant\theta\leqslant\pi/2$.
When $\pi/2<\theta\leqslant\pi$, $\delta N\leqslant 0$ (i.e.,
$N'_+\leqslant N'$) for all $d$. Figure 7 shows that $\delta N$ can
be indeed larger than zero, i.e., $N'_+(\rho_+)>N'(\varrho)$, in a
large region of $d$ and $\theta$. More importantly, when $d$ is very
large and close to one, meaning the quantum channels are very noisy
and the coherence of the transmitted particles degenerate heavily,
$N'_+(\rho_+)$ can overstep $N'(\varrho)$ in almost all the range
$0<\theta<\pi/2$. As a matter of fact, the larger $d$ is, the larger
range of $\theta$ is allowed to be selected for ensuring $\delta
N>0$. It implies that the larger $d$ is, the more flexible the ADS
is. Moreover, if we take a measurement angle $\theta'$ that is
slightly less than $\pi/2$, $N'_+(\rho_+)$ is nearly always larger
than $N'(\varrho)$.

\begin{figure}
\includegraphics[width=8cm,height=6cm]{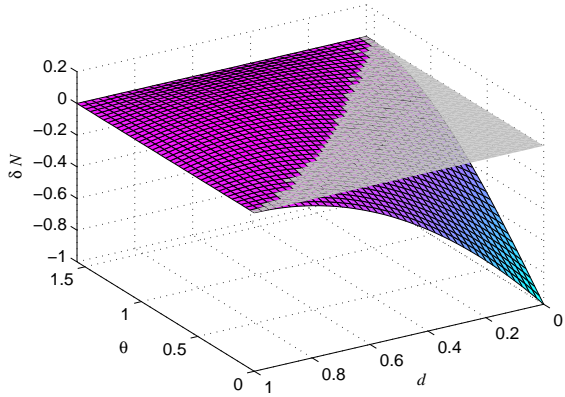}
\caption{(Color online) $\delta N$ as a function of $d$ and
$\theta$, where $\theta$ ranges from 0 to $\pi/2$.}
\end{figure}

As to $d_3>0$, we only consider $d_3$ is very small relative to $d$,
due to the fact that qubit 3 is not transmitted remotely. That is to
say, the ratio of $d_3$ to $d$ is far less than unit. On the other
hand, it has been pointed out that if one selects a measurement
angle $\theta'$ which is close to but less than $\pi/2$,
$N'_+(\rho_+)$ is larger than $N'(\varrho)$ for almost the whole
regime of $0<d<1$. Based on these considerations, we plot $\delta N$
as a function of $d$ and $r=d_3/d$ in Fig.~8 with
$\theta\equiv\theta'=1.5$ and $0\leqslant r\leqslant 0.1$. It can be
seen that even when $d_3$ takes nonzero values, $N'_+(\rho_+)$ can
be larger than $N'(\varrho)$ for almost all values of $d$. It is
worth pointing out that the increase in $d_3$ will lead to the
increase in the probability $P_+$ of obtaining the state
$\rho_+^{(12)}$ for a fixed $\theta$, because $P_+$ is proportional
to the product of $d_3$ and $\cos\theta$ as given in Eq.~(\ref{P+}).
Now we can safely conclude that the aforementioned ADS is able to
enhance, with a certain probability, the quality of bipartite
entanglement distribution, compared to DDS in the above-mentioned
case.

\begin{figure}
\includegraphics[width=8cm,height=6cm]{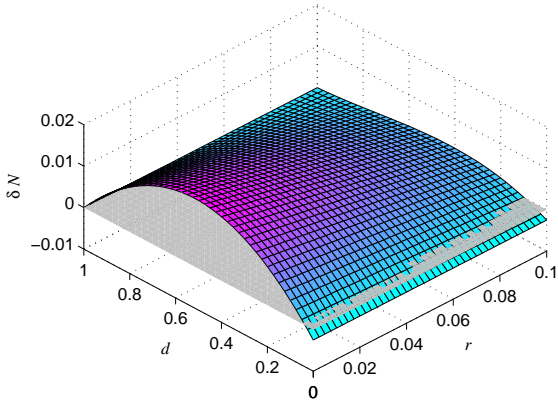}\\
\caption{(Color online) $\delta N$ versus $d$ and $r$ ($=d_3/d$),
where $r\in[0,0.1]$ and $\theta\equiv\theta'=1/5$.}
\end{figure}

\section{Concluding remarks}
In summary, we have investigated the effect of quantum decoherence
on the localization of a three-qubit GHZ state to a two-qubit
entangled state. We used two different entanglement measures,
negativity and FEF, to quantify the resulting bipartite entanglement
after localization procedure. It turns out that the optimal
measurement basis in the noise-free case is no more the optimal one
under the amplitude noise. Moreover, the depolarizing noise has
different influence from amplitude noise on the entanglement
localization. The difference of the effects and the change of the
optimal measurement bases justify the necessity of investigating the
entanglement localization in various noisy environments. It has also
been shown that the optimal measurement basis in the concept of
localizable-entanglement does not match to the one for optimizing
the practical applications of entanglement localization.
Furthermore, we found that the idea of entanglement localizing could
be used to probabilistically improve the equality of bipartite
entanglement distribution. These findings shed new insights into
entanglement manipulations and transformations, and provide a new
idea of entanglement distributing against decoherence as well.

Although the results above are obtained from the case that the
initial multipartite entangled resource is a three-qubit GHZ state,
the conclusions could be directly generalized to the case involving
$N$-qubit ($N>3$) GHZ states. It is deserved to research the effects
of different types of quantum noises on entanglement localization
and distribution for a variety of multipartite entangled states.

\begin{acknowledgements}
This work was supported by the China Postdoctoral Science Foundation
funded project (Grant No.~2013T60769 and No.~2012M511729), the NSFC
(Grant No.~11004050 and No.~11375060), the 973 Program (Grant No.
2013CB921804), the Hunan Provincial Natural Science Foundation
(Grant No.~2015JJ3029), the Hunan Provincial Applied Basic Research
Base of Optoelectronic Information Technology (Grant No.~GDXX007),
and the construct program of the key discipline in Hunan province.
\end{acknowledgements}

\end{document}